\documentstyle{lamuphys}
\makeatletter
\let\chapter\hid@chapter
\makeatother
\input{psfig.tex}
\begin{document}
\pagenumbering{arabic}
\title{The Large Scale X-ray Emission from M87}

\author{D. E. Harris\inst{1}, J. A. Biretta\inst{2}, and
W. Junor\inst{3}}

\institute{Smithsonian Astrophysical Observatory, 60 Garden St,
Cambridge MA 02138, USA
\and
Space Telescope Science Institute, 3700 San Martin Dr., Baltimore, MD,
21218 USA
\and
University of New Mexico, 800 Yale Blvd. NE, Albuquerque, NM, 87131 USA}

\maketitle

\begin{abstract}
We describe asymmetrical features in a long exposure X-ray
map\index{X-ray map} of M87 made with the ROSAT\index{ROSAT} High
Resolution Imager (HRI).  A bright triangular region is marked by a
linear `spur' along one edge.  The structure of this spur suggests an
interpretation of a tangential view of a shock\index{shock} front 18
kpc long.  None of the brighter features are spatially coincident with
radio or optical structures so we concur with earlier investigators
that most of the emission arises from thermal processes.
\end{abstract}
\section{Introduction}

In addition to the strong X-ray emission (roughly circularly
symmetric) from the gas associated with M87 and the Virgo cluster,
asymmetric X-ray features have been known since the Einstein
Observatory\index{Einstein Observatory} observations were obtained 15
years ago (\cite{ethan}, \cite{feigel}).  From these data, it was
noted that the large scale X-ray emission which is asymmetric was
roughly correlated with the intermediate scale radio structure, and
both inverse Compton emission and thermal bremsstrahlung emission were
considered as possible origins.  More recently it has become clear
that there is not a tight spatial correlation between radio and X-ray
emissions (\cite{boh}), and thus the inverse Compton process is not
the major contributor to the observed X-rays.

In this paper, we present further information on these structures
derived from an analysis of a ROSAT HRI map consisting of many
monitoring observations.  The effective integration time for this map
is 202 ks and we are thus able to delineate weaker features with
relatively high resolution.  The X-ray emissions from the core and
knot A in the radio jet will not be dealt with here (see Harris,
Biretta, and Junor, this volume, for the latest variability data on
the core and knot A).

\section{Construction of the Image}

The inherent spatial resolution of the ROSAT HRI is close to
5$^{\prime\prime}$.  However, the effective resolution can be degraded
by two types of aspect (i.e. pointing) problems: the star trackers
occasionally make gross errors of order 10$^{\prime\prime}$ and there
is often a residual error associated with the spacecraft wobble.  The
first type of error results in erroneous locations in celestial
coordinates and is seen most often when combining observations taken 6
months apart, although occasionally there will be an aberrant
observation interval in the typical observation consisting of 10
or more intervals.  The second type of error produces an ellipsoidal point
response function (PRF) which can smear the 5$^{\prime\prime}$
inherent PRF to something like 7$^{\prime\prime}$ by
10$^{\prime\prime}$.  Since the current objective is to study the
larger scale structures, we have not addressed these errors except for
the alignment of observations.

\begin{figure}
%\vspace{5in}
\centerline{
\psfig{figure=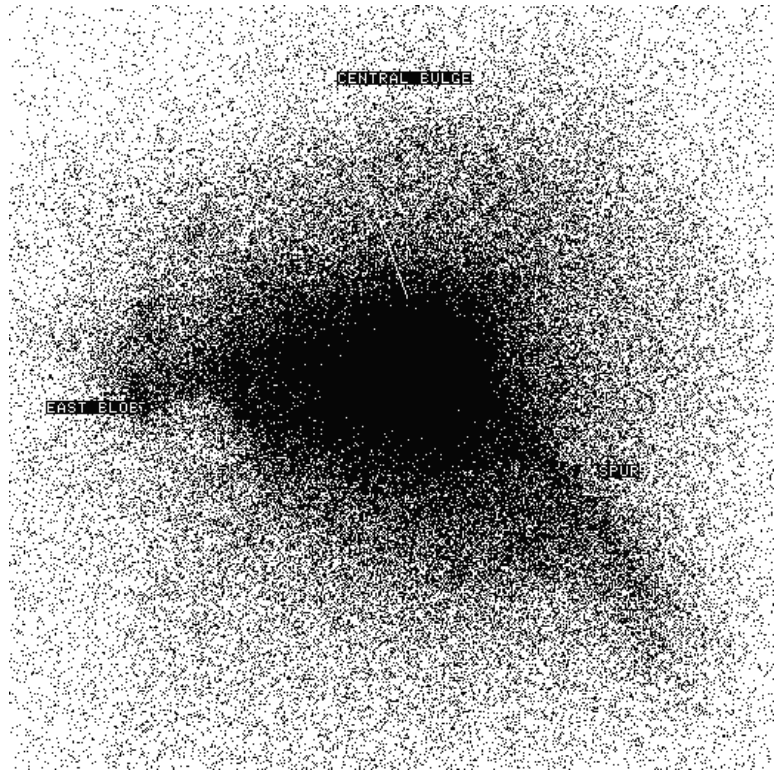}
}
\caption{\label{fig1}The stacked image of M87 from 6 observations.
Each side of the figure has a length of 256 arcsec.  The core and knot
A are within the burned out central region.}
\end{figure}

Our monitoring program has resulted in the accumulation of 5
observations ranging from 30 to 44 ksec each, at 6 month intervals
between 1995Jun and 1997Jun.  To these we added the 13.9 ks HRI
observation obtained in 1992Jun, resulting in a total of 202 ks.
Before adding each observation, we measured the position of both the
core and knot A, and shifted each observation (up to
6$^{\prime\prime}$) so as to align all the data to the reference frame
of the 1992Jun observation.  The resulting events file is represented
by the map in Fig.~\ref{fig1}, to which we have added labels for some
of the features described in section \ref{features}.

\begin{figure}
\centerline{
\psfig{figure=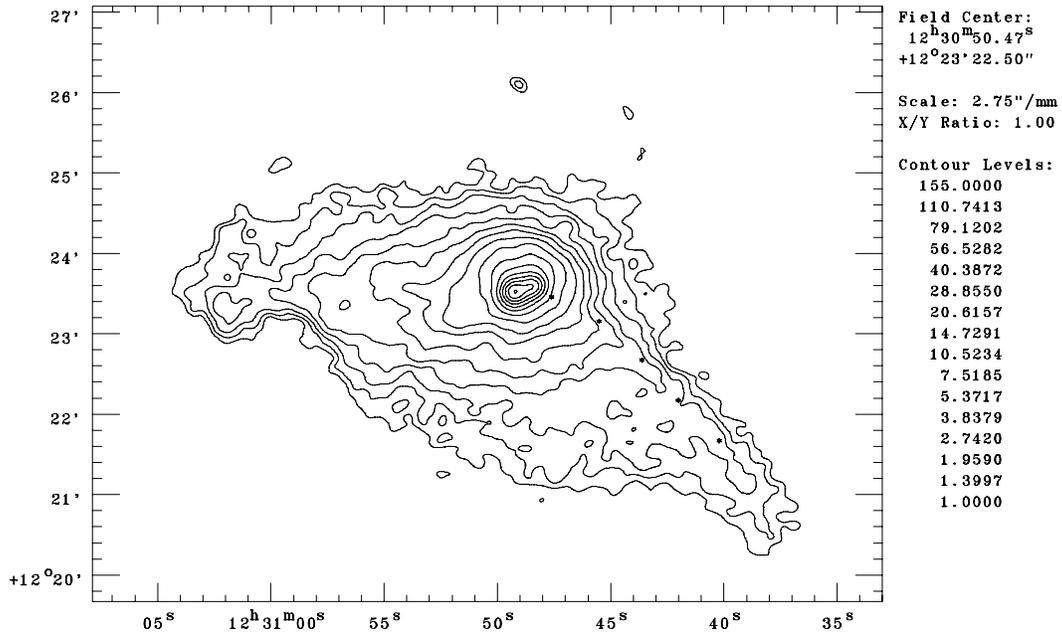,width=15cm}
}
\caption{\label{fig2}The residual X-ray map after subtraction of the
power law model.  Contour intervals are logarithmic, starting at one
count per pixel and ending at 155 counts per pixel.  The pixel size is
one arcsec.  The map has been smoothed with a Gaussian of FWHM =
8$^{\prime\prime}$.  The asterisks indicate measured positions along
the ridge line of the SW spur.}
\end{figure}

Since most of the asymmetrical structure occurs to the East and SW of
the core, we made a radial profile only for the hemisphere between
PA=240$^{\circ}$ and 60$^{\circ}$.  We chose the center for the radial
profile to be $\approx$ 7$^{\prime\prime}$ north of the core since
that position appeared to be the center of the quasi circular `bulge'
of high brightness surrounding the core and knot A.  We made no effort
to fit this central bulge (radius $\approx$ 40$^{\prime\prime}$) but
found that a power law with slope of 1.10 was a reasonable fit to the
brightness between 40$^{\prime\prime}$ and 900$^{\prime\prime}$.  We
then subtracted this power law ({1+(r/a)}$^{-1.1}$) from the image,
adjusting the normalization until there were few negative regions.
The results are shown in Fig. \ref{fig2}.

\section{Description of Discrete Features}\label{features}

For the most part, the following sections will describe some of the
more obvious features of the X-ray distribution without a quantitative
analysis.  The latter subject will be addressed elsewhere.

\subsection{SW spur}

\begin{figure}
\centerline{
\psfig{figure=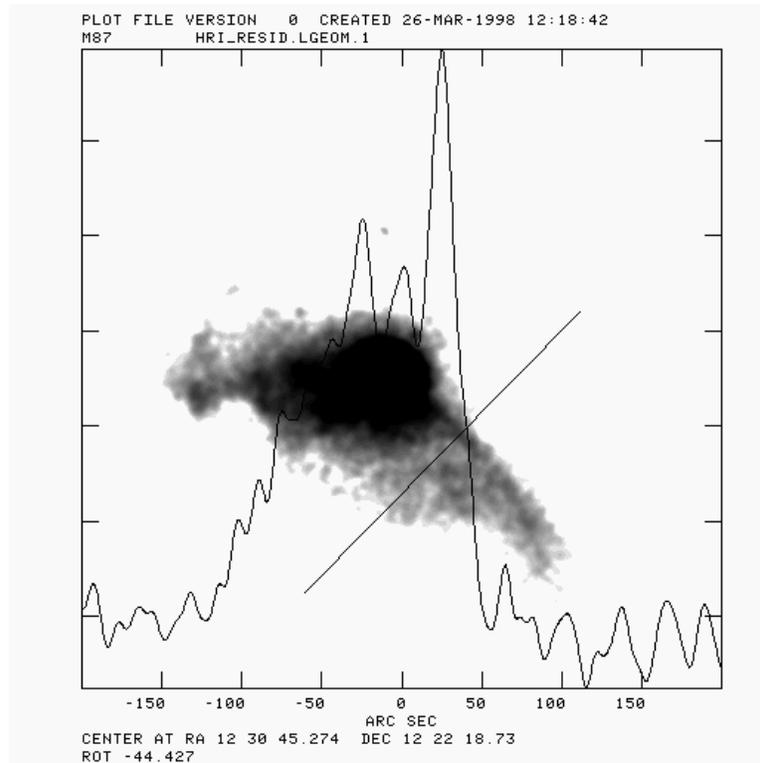}
}
\caption{\label{fig3}A profile though the X-ray
spur\index{X-ray!spur}.  Note the sharp leading edge on the right
side.  The X-ray map has had the model power law subtracted and was
smoothed with a Gaussian of FWHM = 8$^{\prime\prime}$.  The straight
line shows the location of the measured profile. }
\end{figure}

The most striking X-ray feature evident in Figs.~\ref{fig1}
and~\ref{fig2} is the spur\index{X-ray!spur} of emission extending
from the central region $\approx$ 4$^{\prime}$ toward the south west
in a slightly curved trajectory.  For a distance of 16 Mpc to M87,
4$^{\prime}$ corresponds to 19 kpc.  The NW edge is essentially
unresolved but the brightness falls off more gradually behind it
towards the SE.  This is demonstrated in Fig.~\ref{fig3}.

There is no optical or radio feature corresponding to the spur, and
this agrees with the results of \cite{boh} who used a spectral
analysis of the ROSAT Position Sensitive Proportional Counter
observation to conclude that the asymmetric X-ray emission was
thermal, but from a cooler gas than the surrounding cluster emission.
However, from a preliminary hardness ratio\index{X-ray!hardness
ratio} map of our HRI data, there is some indication that the spur is
actually harder than the surrounding emission.
 
As is often the case, it is difficult to reproduce the detail that can
be obtained by visual inspection of various versions of an image.  For
this reason, we have added asterisks at several positions, measured by
eye, of the ridge line of the spur\index{X-ray!spur}
(Fig.~\ref{fig2}).  This aid in tracing the spur closer into the core
allows us to conjecture that the spur may be causally related to knot
A.  A possible explanation for the spur (and perhaps for much of the
Eastern arm) would be that it is a tangential view of a bow
shock\index{shock!bow} associated with the radio jet: either knot A
itself or just beyond where the jet bends sharply to the South.  The
excess X-ray emission which we call the spur would then come from a
relatively long path length through the higher density gas compressed
behind the bow shock.

A quite different scenario, but also involving a shock\index{shock}
front, turns the argument around. If the shock is not caused by the
radio jet, it would still provide a discontinuity in the ambient
density, temperature, and pressure.  It is just such conditions which
can be the causal agent for the genesis of an internal shock in a
radio jet (e.g. \cite{hooda}); i.e. the change in external pressure is
the reason that knot A exists at this location.  For this scenario,
the spur would be the emission from lines of sight which are
tangential to the curved bow shock.  From the scale of the spur and
eastern arm, the most likely cause of such a shock would be the
interaction of the galaxy's ISM\index{ISM!interaction with ICM} with
the ICM\index{ICM}.  Although M87 is not thought to be moving
substantially with respect to the Virgo ICM, \cite{bing} has argued
that a merger\index{merger} between the M87 subgroup and the M86
subgroup is in progress, and this could provide the required relative
velocity between gas distributions to form a weak shock.

The chief criticism of the association between an ICM/ISM shock and
knot A is that one would not expect knot A to be seen, at least in
projection, near the edge of the shock.  This is because the standard
interpretation of the M87 jet is that it is aimed towards us, not much
more than 30$^{\circ}$ to the line of sight.

\subsection{Other Features Comprising the Triangular Emission Region}

The brightest part of the residual image (Fig.~\ref{fig2}) forms a
roughly triangular region on the sky with the spur delineating the NW
edge.  The southern `base' of the triangle is not straight, nor is it
well defined; the emission falls away gradually to the south.  The
brightest part of the triangle aside from the core and knot A is the
central bulge, a quasi circular distribution with a characteristic
radius of 40$^{\prime\prime}$.  Extending towards the East is the
`eastern arm' which has a complex morphology.  The easternmost feature
in Fig.~\ref{fig2} is a quasi circular emission region, `the eastern
blob'.  Note the sharp notch in the base of the triangle near the
eastern end of the base.  It is this notch which defines the eastern
blob, and it is tempting to interpret this notch as the effect of
absorption\index{X-ray!absorption}.  However, there is no
corresponding feature on the hardness ratio map.

\subsection{Lower Brightness Extended Emission}

Referring now to Fig.~\ref{fig1}, we see a lower brightness region to
the north of the triangle.  This region is remarkable in that it has 
a well defined northern boundary which is quasi circular, with a
radius of 3.5$^{\prime}$, centered about 30$^{\prime\prime}$ East of
the core of M87.  The circular boundary originates in the eastern blob
and can be traced for over 90$^{\circ}$.  At a somewhat lower
brightness, one can see extended emission below the `base' of the
triangle; on some displays, it appears to be somewhat striated.
Finally, to the NE of the Eastern Blob, outside of of the circular
boundary, there is a `fan' of lower brightness emission which ends
near the left border of Fig.~\ref{fig1}. 

\section{Acknowledgments}
The work at SAO was partially supported by NASA contract NAS5-30934
and grant NAG5-2960; that at STScI by NASA grant NAG5-2957.

%

%
% ---- Bibliography ----
%

\end{document}